\documentclass[aps,pra,10pt,floatfix]{revtex4}

\usepackage{epsfig}
\usepackage{bm}
\newcommand{\p}{\partial}
\newcommand{\half}{\frac{1}{2}}
\newcommand{\eab}{\epsilon_{\alpha\beta}}
\newcommand{\bnabla}{\bm{\nabla}}

\newcommand{\pontryagin}{{\cal N}}
\newcommand{\Ee}{E_{\rm e}}
\newcommand{\Ea}{E_{\rm a}}
\newcommand{\lex}{\ell_{\rm ex}}

\newcommand{\am}{L}
\newcommand{\bOmega}{\overline{\Omega}}
\newcommand{\bzeta}{\overline{\zeta}}
\newcommand{\qfactor}{q}

\begin{document}

\baselineskip14pt

\title{Dynamics of vortex-antivortex pairs in ferromagnets}
\author{Stavros Komineas$^1$ and Nikos Papanicolaou$^2$}

\affiliation{$^1$Max-Planck Institute for the Physics of Complex Systems,
N\"othnitzer Str. 38, 01187 Dresden, Germany \\
$^2$Department of Physics, and Institute of Plasma Physics, University of Crete,
Heraklion, Greece}

\date{\today}

\begin{abstract}
\baselineskip6pt
\small
We study the dynamics of vortex-antivortex (VA) pairs
in an infinitely thin ferromagnetic film with easy-plane anisotropy.
These are localized excitations with finite energy that are characterized
by a topological (skyrmion) number $\pontryagin = 0,\pm 1$.
Topologically trivial $(\pontryagin=0)$ VA pairs undergo Kelvin motion
analogous to that encountered in fluid dynamics.
In contrast, topologically nontrivial $(\pontryagin = \pm 1)$ VA pairs
perform rotational motion around a fixed guiding center.
We present the results of a detailed study in both cases and further
demonstrate that in the presence of dissipation a rotating $\pontryagin = \pm 1$
 VA pair shrinks to a point and is annihilated, due to the discreteness
of the lattice, thus leading to a ``topologically forbidden''
$\Delta\pontryagin = 1$ process.
We argue that the latter process underlies the experimentally observed
vortex core switching whereby the polarity of a single vortex
is reversed after collision with an $\pontryagin = 0$ VA pair created
by a burst of an applied alternating magnetic field.
\end{abstract}

\pacs{}
\maketitle

\baselineskip12pt

\section{Introduction}
\label{sec:intro}

The best known examples of topological magnetic solitons are magnetic bubbles
or skyrmions observed in abundance in ferromagnetic films with easy-axis
anisotropy \cite{malozemoff,hubert}.
The experimental situation is less clear in the case of ferromagnets with easy-plane
anisotropy. The relevant topological structures are theoretically predicted
to be half skyrmions or vortices, with a logarithmically divergent energy
that may inhibit production of an isolated vortex on an infinite film.
Thus the early studies of magnetic vortex dynamics
had been mostly theoretical \cite{huber82,mertens99}
drawing on various analogies with related work
on ferromagnetic bubbles \cite{thiele74}, 
with vortex dynamics in classical fluids and superfluids
\cite{batchelor,saffman,donnelly},
as well as with the dynamics of interacting electric charges in a uniform
magnetic field.

The situation has changed dramatically in recent years.
It has been realized that a disc-shaped magnetic element, with a diameter
of a few hundred of nanometers, provides an excellent geometry for the
realization of a magnetic vortex configuration.
In particular, the exchange energy is finite on a finite element while the
magnetostatic field vanishes everywhere except at the vortex core.
As a result, the vortex is actually the lowest energy magnetic state in a
disc-shaped element. In other words, interest in the vortex stems from
the fact that it is a nontrivial magnetic state which can, nevertheless,
be spontaneously created in magnetic elements \cite{raabe00}.

It is then natural to ask whether nontrivial magnetic states other than
the single vortex may play an important role in the dynamics
of magnetic elements \cite{shigeto02,castano03}.
An answer to this question comes from a somewhat unlikely direction.
Recent experiments have shown a peculiar dynamical behavior of vortices
and magnetic domain walls when these are probed by external magnetic fields.
Vortices may switch their polarity under the influence of a very weak
external magnetic field of the order of a few mT \cite{neudert05,waeyenberge06}.
The same switching phenomenon was observed by passing an a.c. electrical current
through a magnetic disc \cite{yamada07}.
Since the polarity of the vortex contributes to its topological structure,
the switching process clearly implies a discontinuous (topologically forbidden)
change of the magnetic configuration.
This is certainly a surprise especially because the external field is rather weak.
The key to this phenomenon is the appearance of vortex pairs which are
spontaneously created in the vicinity of existing vortices \cite{waeyenberge06,hertel07}.
The creation of topological excitations (vortex pairs) by alternating external fields
had been anticipated by an early study based on collective coordinates
\cite{pokrovskii85}.

In this paper we study vortex-antivortex pairs (VA pairs) which are nontrivial
magnetic states that play an important role in the dynamics
of magnetic elements.
However, unlike a single vortex, a VA pair is a localized object whose energy
remains finite even on an infinite film. It is then reasonable to expect that
the essential features of the dynamics of VA pairs can be understood in
the infinite-film approximation which is adopted in the following.
A brief summary of the relevant dynamical equations and related topological
structures is given in Section~\ref{sec:model}.
In Section~\ref{sec:kelvin} we study a VA pair in which the vortex and the antivortex
carry the same polarity. Such a pair is shown to undergo translational
Kelvin motion analogous to that observed in fluid dynamics \cite{papanicolaou99}.
In Section~\ref{sec:rotate} we study a VA pair in which the vortex and the antivortex
carry opposite polarities. Such a pair is shown to behave as a rotating
vortex dipole \cite{komineas07} because its topological structure
is substantially different than that of the pair in Kelvin motion.

The preceding results are combined in Section~\ref{sec:switching}
to demonstrate that a rotating vortex dipole may be annihilated by a quasi-continuous
process in spite of its nontrivial topological structure.
In particular, no energy barrier has to be overcome in contrast to the usual
expectations for topological solitons.
This opens the possibility for switching mechanisms between topologically
distinct states in ferromagnets.
The possibility to change the topological structure leads to a dramatic change
in the magnetization dynamics as a VA pair is created or annihilated.
Such a pair annihilation process lies in the heart of the counter-intuitive
vortex polarity switching event that was observed in magnetic elements
\cite{neudert05,waeyenberge06,yamada07}.

Some concluding remarks are summarized in Section~\ref{sec:conclusion}.
Finally, an Appendix is devoted to a brief description of the dynamics
of interacting electric charges in a uniform magnetic field,
which also exhibits most of the peculiar features of the dynamics of VA pairs.

\section{The model}
\label{sec:model}

A ferromagnet is characterized by the magnetization $\bm{m} = (m_1, m_2, m_3)$
measured in units of the constant saturation magnetization $M_s$.
Hence $\bm{m}$ is a vector field of unit length, $\bm{m}^2 = m_1^2+m_2^2+m_3^2 = 1$,
but is otherwise a nontrivial function of position and time
$\bm{m} = \bm{m}(\bm{r},t)$ that satisfies the rationalized Landau-Lifshitz (LL)
equation
\begin{equation}  \label{eq:lle}
\frac{\p\bm{m}}{\p t} = \bm{m} \times \bm{f}, \quad
\bm{f} \equiv \Delta\bm{m} - \qfactor\, m_3\, \bm{\hat{e}}_{\rm 3}, \quad
\bm{m}^2 = 1.
\end{equation}
Here distances are measured in units of the exchange length
$\lex = \sqrt{A/2\pi M_s^2}$, where $A$ is the exchange constant,
and the unit of time is $\tau_0 \equiv 1/(4\pi\gamma M_s)$ where $\gamma$ is the
gyromagnetic ratio.
Typical values are $\lex \sim 5{\rm nm}$
and $\tau_0 \sim  10{\rm ps}$ which set the scales for the phenomena
described by Eq.~(\ref{eq:lle}).
To complete the description of the LL equation we note that we consider
ferromagnetic materials with uniaxial anisotropy.
Then $\bm{\hat{e}}_{\rm 3}$ in Eq.~(\ref{eq:lle}) is a unit vector along
the symmetry axis and the dimensionless parameter $\qfactor \equiv K/2\pi M_s^2$,
where $K$ is an anisotropy constant, measures the strength of anisotropy.
In particular, $\qfactor$ is taken to be positive throughout this paper,
a choice that corresponds to easy-plane ferromagnets.

An important omission in Eq.~(\ref{eq:lle}) is the demagnetizing field produced
by the magnetization itself \cite{malozemoff,hubert}. However, in the limit
of a very thin film, the effect of the demagnetizing field is thought to
amount to a simple additive renormalization of the anisotropy constant
\cite{gioia97}.
Also note that we may perform the rescalings $\sqrt{\qfactor} \bm{r} \to \bm{r}$ and
$\qfactor t \to t$ which further renormalize the units of space and time
discussed earlier and lead to a completely rationalized LL equation
where we may set $\qfactor=1$ without loss of generality.
With this understanding, all calculations presented in this paper are based
on a two-dimensional (2D) restriction of Eq.~(\ref{eq:lle}),
i.e., $\bm{r} = (x,y)$ and $\Delta = \p^2/\p x^2 + \p^2/\p y^2$, while
$\qfactor$ is set equal to unity without further notice.

The effective field $\bm{f}$ in Eq.~(\ref{eq:lle}) may be derived from
a variational argument:
\begin{equation}  \label{eq:energy}
\bm{f} = -\frac{\delta E}{\delta \bm{m}}\,; \quad
E = \half \int{\left[ (\bnabla\bm{m})^2 + m_3^2 \right]\, dx dy}
\end{equation}
where $E$ is the conserved energy functional.
A standard Hamiltonian form is obtained by resolving the constraint $\bm{m}^2=1$
through, say, the spherical parameterization
\begin{equation}  \label{eq:thetaphi}
m_1 + i\, m_2 = \sin\Theta\;e^{i\Phi}, \quad m_3 = \cos\Theta.
\end{equation}
The LL equation is then written as
\begin{equation}  \label{eq:llesymplectic}
\frac{\p\Phi}{\p t} = \frac{\delta E}{\delta \Pi}\,, \quad 
\frac{\p\Pi}{\p t} = -\frac{\delta E}{\delta \Phi}
\end{equation}
where $\Pi\equiv \cos\Theta$ is the canonical momentum conjugate to the
azimuthal angle $\Phi$. Taking into account the specific form
of the energy in Eq.~(\ref{eq:energy}) or
\begin{equation}  \label{eq:energy1}
E = \half \int \left[ (\bnabla\Theta)^2 + \sin^2\Theta\, (\bnabla\Phi)^2
    + \cos^2\Theta\right]\, dx dy
\end{equation}
the Hamilton equations (\ref{eq:thetaphi}) yield
\begin{eqnarray}  \label{eq:llethetaphi}
\sin\Theta\, \frac{\p \Phi}{\p t} & = &
   \Delta\Theta + [1 - (\bnabla\Phi)^2]\,\cos\Theta\,\sin\Theta\,, \nonumber \\
\sin\Theta\, \frac{\p \Theta}{\p t} & = &
                  - \bnabla\cdot (\sin^2\Theta\, \bnabla\Phi).
\end{eqnarray}
Another useful parameterization is obtained through the stereographic variable $\Omega$:
\begin{equation}  \label{eq:omega}
\Omega = \frac{m_1 + i\,m_2}{1+m_3}\,;\quad
m_1 + i\,m_2 = \frac{2\Omega}{1+\bOmega\Omega}\,,\quad
m_3 = \frac{1-\bOmega\Omega}{1+\bOmega\Omega}
\end{equation}
in terms of which the LL equation reads
\begin{equation}  \label{eq:lleomega}
i\,\frac{\p \Omega}{\p t} + \Delta\Omega + \frac{1-\bOmega\Omega}{1+\bOmega\Omega}\,\Omega =
\frac{2\bOmega}{1+\bOmega\Omega}\,(\bnabla\Omega\cdot\bnabla\Omega)
\end{equation}
where $\bOmega$ is the complex conjugate of $\Omega$.
Equations (\ref{eq:lle}),(\ref{eq:llethetaphi}), and (\ref{eq:lleomega})
are three equivalent versions of the LL equation and may be used at convenience
depending on the specific calculation considered.

A key quantity for describing both topological and dynamical properties of the 2D
LL equation is the local topological vorticity $\gamma = \gamma(x,y,t)$ defined
from \cite{papanicolaou91,komineas96}:
\begin{equation}  \label{eq:vorticity1}
\gamma = \eab\, \p_\alpha\Pi\p_\beta\Phi
       = \eab\, \sin\Theta\, \p_\beta\Theta \p_\alpha\Phi
       = \half \eab\, (\p_\beta\bm{m} \times \p_\alpha\bm{m}) \cdot \bm{m},
\end{equation}
where the usual summation convention is invoked for the repeated indices $\alpha$
and $\beta$, which take over two distinct values corresponding to the two spatial
coordinates $x$ and $y$, and $\eab$ is the 2D antisymmetric tensor.
In particular, one may consider the total topological vorticity $\Gamma$
and the Pontryagin index or skyrmion number $\pontryagin$ defined from
\begin{equation}  \label{eq:skyrmionnumber}
\Gamma = \int{\gamma\, dx dy},\quad  \pontryagin = \frac{\Gamma}{4 \pi}.
\end{equation}
A naive partial integration using Eq.~(\ref{eq:vorticity1})
yields $\Gamma=0=\pontryagin$ for all magnetic configurations for which
such an integration is permissible.
However, nonvanishing values for $\Gamma$ and $\pontryagin$ are possible
and are topologically quantized.
Specifically, for field configurations that approach a constant (uniform)
magnetization at spatial infinity, the skyrmion number $\pontryagin$ is quantized
according to $\pontryagin = 0,\pm 1,\pm 2, \ldots$. Half integer values are also
possible in the case of field configurations with more complicated structure
at infinity such as half skyrmions or vortices (see below).

The local topological vorticity $\gamma$ is also important for an unambiguous
definition of conservation laws in the LL equation.
Hence the linear momentum (impulse) $\bm{P} = (P_x, P_y)$ is defined from
\begin{equation}  \label{eq:linmomentum}
P_x = - \int{y\gamma\, dx dy},\quad P_y = \int{x\gamma\, dx dy},
\end{equation}
while the angular momentum (impulse) is given by
\begin{equation}  \label{eq:angmomentum}
\am = \half\int{\rho^2\gamma\, dx dy}
\end{equation}
where $\rho^2 = x^2+y^2$. Since detailed discussions of these conservation laws have 
already appeared in the literature \cite{papanicolaou99,papanicolaou91,komineas96}
we simply note here that analogous conservation laws were defined as moments
of ordinary vorticity in fluid dynamics \cite{batchelor,saffman}.

\begin{figure}
\epsfig{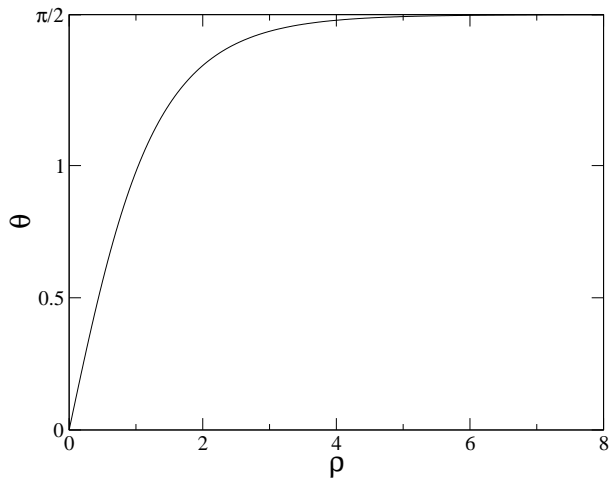}
   \caption{Profile of a single vortex calculated numerically.
The complete solution is given by Eq.~(\ref{eq:vortex}).
   }
 \label{fig:vortex}
\end{figure}

We first search for static (time independent) solutions of the LL equation
which may be obtained by omitting time derivatives in Eq.~(\ref{eq:llethetaphi})
and further introducing the axially symmetric ansatz $\Theta=\theta(\rho)$
and $\Phi= \kappa(\phi-\phi_0)$,
where $\rho$ and $\phi$ are the usual cylindrical coordinates
$(x =\rho\cos\phi,\, y =\rho\sin\phi)$, $\kappa=\pm 1$
will be referred to as the vortex number,
and $\phi_0$ is an arbitrary constant phase reflecting the
azimuthal invariance. The resulting ordinary differential equation
for the amplitude $\theta=\theta(\rho)$ is solved numerically
with standard boundary condition $\theta(\rho \to \infty) = \pi/2$ and the result
is shown in Fig.~\ref{fig:vortex}.
The corresponding magnetization is then given by
\begin{equation}  \label{eq:vortex}
m_1 + i\, m_2 = \sin\theta\, e^{i\kappa(\phi-\phi_0)},
\quad m_3 = \lambda\cos\theta,
\end{equation}
where $\lambda=\pm 1$ will be called the polarity.
The total energy is accordingly reduced to
\begin{equation}  \label{eq:energy2}
E = \half \int\limits_0^\infty\left[\left(\frac{\p\theta}{\p\rho}\right)^2
 + \frac{\sin^2\theta}{\rho^2} + \cos^2\theta \right]\, (2\pi\rho d\rho).
\end{equation}
where the centrifugal (second) term  is logarithmically divergent
for the assumed boundary condition at spatial infinity.
However, the anisotropy energy given by the last term is finite and is
actually predicted to be
\begin{equation}  \label{eq:virial1}
\Ea = \half \int\limits_0^\infty \cos^2\theta\, (2\pi\rho d\rho) = \frac{\pi}{2}
\end{equation}
by a careful derivation of a suitable virial relation \cite{komineas98}.
Finally, the total vorticity $\Gamma$ and the skyrmion number $\pontryagin$
are calculated from Eq.~(\ref{eq:skyrmionnumber}) to be
\begin{equation}  \label{eq:vortextopo}
\Gamma = -2\pi\kappa\lambda\,,\quad \pontryagin = -\half\,\kappa\lambda\,,
\end{equation}
where the vortex number $\kappa=\pm 1$ and the polarity $\lambda=\pm 1$
may be taken in any combination.
Thus, we must consider four possibilities; namely, a vortex that comes
in two varieties ($\kappa=1, \lambda=\pm 1$) and thus $\pontryagin= \mp 1/2$,
and an antivortex which also comes 
in two varieties ($\kappa=-1, \lambda=\pm 1$) and thus $\pontryagin= \pm 1/2$.
In all cases the calculated static solution is a topological soliton
with half integer skyrmion number, in contrast to ordinary skyrmions
(such as magnetic bubbles) which carry integer $\pontryagin$.

Our main aim in the continuation of this paper is to search for nontrivial
solutions that may combine a vortex and an antivortex (VA pair)
in a way that the total energy is finite.
We leave aside for the moment the LL equation and construct model VA pairs
in terms of the basic single-vortex configuration described above.
This is easily accomplished by invoking the stereographic variable $\Omega$
of Eq.~(\ref{eq:omega}) to write a single ($\kappa,\lambda$) vortex located at the
origin of coordinates as
\begin{equation}  \label{eq:omega-thetaphi}
\Omega = \frac{\sin\theta}{1+\lambda \cos\theta}\, e^{i\kappa(\phi-\phi_0)}
\end{equation}
where $\theta=\theta(\rho)$ is the vortex profile taken from Fig.~\ref{fig:vortex}
or simply the model profile defined from $\cos\theta=1/\cosh\rho$ and
$\sin\theta=\tanh\rho$.
We now produce two replicas of the basic vortex to describe a pair of vortices
by the product ansatz
\begin{equation}  \label{eq:productomega}
\Omega = \Omega_1\,\Omega_2
\end{equation}
where $\Omega_1$ is configuration (\ref{eq:omega-thetaphi}) applied for
$(\kappa,\lambda) = (\kappa_1,\lambda_1)$ and the origin displaced to, say,
$(x,y) = (-d/2,0)$
while $\Omega_2$ is a $(\kappa_2,\lambda_2)$ vortex located around $(x,y) = (d/2,0)$.
The skyrmion number of this configuration is given by
\begin{equation}  \label{eq:addS}
\pontryagin = -\half\, (\kappa_1\lambda_1 + \kappa_2\lambda_2)
\end{equation}
and its energy is finite only if we restrict attention to vortex-antivortex (VA) pairs
$(\kappa_1=-\kappa_2)$.
For definiteness we choose $\kappa_1=1$ and $\kappa_2=-1$ and thus the skyrmion number
\begin{equation}  \label{eq:subtractS}
\pontryagin = -\half\, (\lambda_1-\lambda_2)
\end{equation}
depends on the polarities $\lambda_1$ and $\lambda_2$.
Now a VA pair with equal polarities $(\lambda_1=\lambda_2 = \pm 1)$ is topologically
trivial $(\pontryagin = 0)$. In contrast, a VA pair with opposite polarities
is topologically equivalent to a skyrmion
($\pontryagin = 1$, for $\lambda_1 = -1$ and $\lambda_2=1$)
or an antiskyrmion ($\pontryagin = -1$, for $\lambda_1 = 1$ and $\lambda_2=-1$).
In all three cases configuration (\ref{eq:productomega}) carries finite energy
because its overall phase cancels out at spatial infinity where the magnetization
approaches the uniform configuration $\bm{m}=(1,0,0)$
modulo an overall azimuthal rotation which depends on the choice
of individual  phases $\phi_0$ in the ansatz (\ref{eq:productomega}).
This explains, in particular, why VA pairs are characterized by an integer
skyrmion number ($\pontryagin=0,\pm 1$).

Needless to say, the VA pairs constructed above are not solutions of the
LL equation. Yet an interesting picture arises when configuration (\ref{eq:productomega})
is used as an initial condition in the complete equation (\ref{eq:lleomega}).
A topologically trivial ($\pontryagin=0$) VA pair undergoes Kelvin motion
in which the vortex and the antivortex initially located at a relative distance $d$
along the x axis move in parallel along the y axis with nearly constant velocity.
In contrast, a topologically nontrivial ($\pontryagin=\pm 1$) VA pair
undergoes rotational motion around a fixed guiding center at nearly constant
angular velocity. In both cases the main trajectories are decorated by Larmor-type
oscillations \cite{volker94} which are tamed when the relative distance between the
vortex and the antivortex is large.
The effect of the polarity of vortex pairs has been studied experimentally
in patterned ferromagnetic ellipses \cite{buchanan05}. Two vortices were created an ellipse 
and it was found that their dynamics depended on their relative polarity,
as is indicated by Eq.~(\ref{eq:subtractS}).

In the following two sections we examine the two cases in turn.
In particular, we aim at constructing true steady-state solitary waves that
describe VA pairs in pure Kelvin motion for $\pontryagin=0$ and pure rotational motion
for $\pontryagin=\pm 1$.

\section{Kelvin motion}
\label{sec:kelvin}

As the title of this section suggests, VA pairs in Kelvin motion were
originally studied in the context of ordinary fluid dynamics \cite{batchelor,saffman}.
A further analogy exists with the 2D motion of an electron-positron pair
interacting via the Coulomb potential and placed in a uniform magnetic field
perpendicular to the plane.
If the electron and the positron are initially at rest their guiding centers
will move along two parallel straight lines while the actual trajectories
will display the familiar Larmor oscillations. However, when the electron and the
positron are given a common initial velocity such that the Coulomb force is
exactly balanced by the magnetic force,
both the guiding centers and the actual positions of the
charges will move steadily along parallel lines, even though the two sets
of trajectories do not coincide.
The resulting special configuration may be thought of as a peculiar
electron-positron bound state in steady translational motion (see our Appendix).

It is thus reasonable to expect that a solitary wave exists in a 2D easy-plane
ferromagnet which describes a VA pair that proceeds rigidly (without Larmor oscillations)
in a direction perpendicular to the line connecting the vortex and the antivortex,
probably because the mutual force is exactly balanced by a topological
``Magnus force''.
The actual existence of such a solitary wave was established in Ref.~\cite{papanicolaou99}
whose main result is briefly reviewed in the remainder of this section.

As it turns out, when the relative distance is large, the sought after solitary wave
resembles in its gross features  the model VA pair of Eq.~(\ref{eq:productomega})
applied for, say, $\kappa_1=-\kappa_2=1$ and $\lambda_1=\lambda_2=1$
(thus $\pontryagin=0$). We may then invoke this model to motivate some important
asymptotic results valid for large $d$.
For instance, the local topological vorticity $\gamma$ is then peaked around
the two points $(x,y)=(-d/2,0)$ and $(x,y)=(d/2,0)$ with weights $-2\pi$ and $2\pi$
respectively, corresponding to the total vorticities $\Gamma$ of the individual
vortex and antivortex.
Then the impulse defined from Eq.~(\ref{eq:linmomentum}) yields $P_x=0$,
thanks to reflexion symmetry, while $P = P_y \sim 2\pi d$. One may also
invoke the Derrick-like scaling relation applied to the extended energy functional
$F=E- v P$:
\begin{equation}  \label{eq:derrick1}
v P = \int m_3^2\, dx dy = 2\,\Ea
\end{equation}
where $\Ea$ is the total anisotropy energy of the VA pair.
For large $d$ this energy approaches the sum of the anisotropy energies
of the individual  vortex and antivortex, each given by $\pi/2$ in view of the
virial relation (\ref{eq:virial1}). Hence,
$\int{m_3^2\, dx dy} \sim 2 (\pi/2+\pi/2) = 2\pi$ and $vP \sim 2\pi$.
To summarize,
\begin{equation}  \label{eq:estimates1}
P \sim  2\pi d, \quad v P \sim 2\pi, \quad v \sim\frac{1}{d}.
\end{equation}
One may further consider the familiar group-velocity relation
\begin{equation}
v = \frac{dE}{dP}
\end{equation}
in which we insert the estimate $v\sim 2\pi/P$ to obtain an elementary
differential equation for $E$ whose integral is
\begin{equation}  \label{eq:kelvinpairenergy}
E \approx 2\pi\,\ln(P/P_0)
\end{equation}
where $P_0$ is an integration constant that cannot be fixed by the present leading-order
argument. Nevertheless, Eq.~(\ref{eq:kelvinpairenergy}) provides the essence
of the energy-momentum dispersion at large relative distance $d$
or small velocity $v\sim 1/d$ and hence large momentum $P\sim 2\pi d$.

To obtain an accurate numerical solution it is convenient to work
with the stereographic variable $\Omega$ of Eq.~(\ref{eq:omega}).
Then the LL Eq.~(\ref{eq:lleomega}) restricted to a solitary wave in
rigid motion with constant velocity $v$ along, say, the y axis reads
\begin{equation}  \label{eq:lleomega1}
-iv\,\frac{\p \Omega}{\p y} + \Delta\Omega
  + \frac{1-\bOmega\Omega}{1+\bOmega\Omega}\,\Omega =
\frac{2\bOmega}{1+\bOmega\Omega}\,(\bnabla\Omega\cdot\bnabla\Omega)
\end{equation}
and is supplemented by the boundary condition $\Omega \to 1$ at spatial infinity
where the magnetization approaches a uniform configuration.
Once a solution $\Omega = \Omega(x,y;v)$ of Eq.~(\ref{eq:lleomega1}) is obtained
for a specific value of the velocity $v$, the sought after solitary wave
is given by $\Omega(x,y-vt;v)$.

\begin{figure}
\epsfig{file=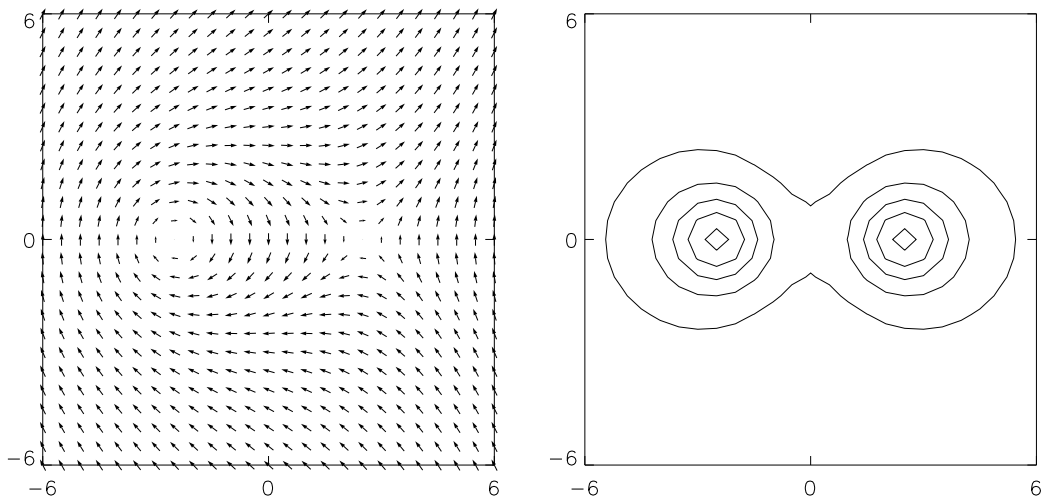,width=0.9\linewidth}
   \caption{Snapshot of a topologically trivial ($\pontryagin=0$)
VA pair in Kelvin motion, illustrated through the ($m_1,m_2$) projection
of the magnetization $\bm{m} = (m_1,m_2,m_3)$ on the plane of the film
(left panel) as well as level contours of $m_3$ (right panel).
The contour levels $m_3=0.1,0.3,0.5,0.7,0.9$ are shown. The outer curve corresponds
to $m_3=0.1$, while the two smallest circles are the contours for $m_3=0.9$
and surround the two vortex centers.
The pair moves along the $y$ axis with velocity $v=0.2$ at a
calculated relative distance $d=4.97$ (along the x axis), energy $E=20$,
and impulse $P=30$.
   }  \label{fig:kelvin02}
\end{figure}

\begin{figure}
\epsfig{file=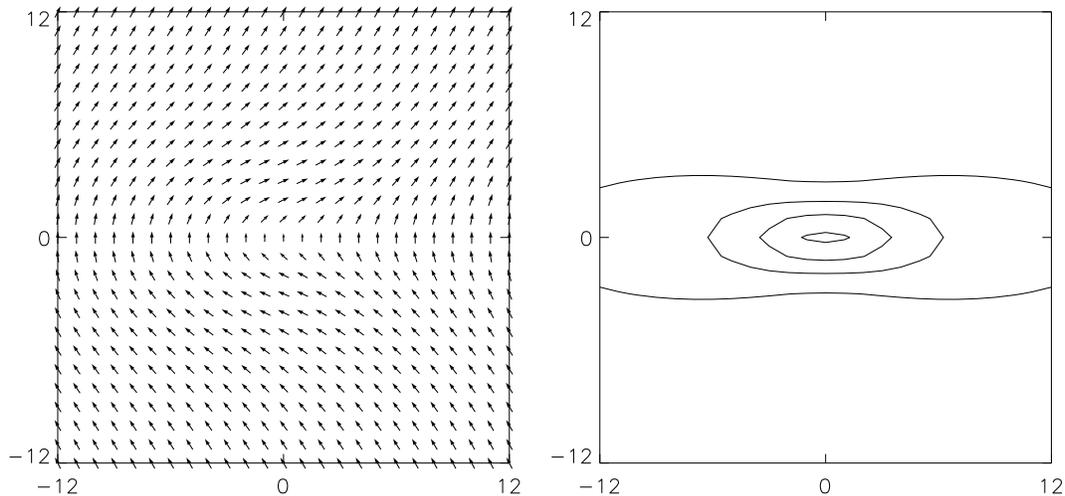,width=0.9\linewidth}
   \caption{Solitary wave in Kelvin motion with $v=0.95$ along
the y axis (conventions as in Fig.~\ref{fig:kelvin02}).
Note that the vortex-antivortex character is lost ($d=0$) and the wave
is a lump with no apparent topological features.
%No contour for $m_3=0.9$ is shown as $m_3$ is everywhere smaller than this value.
The calculated energy and impulse are $E=18$ and $P=17$.
   }  \label{fig:kelvin095}
\end{figure}

Equation (\ref{eq:lleomega1}) was solved numerically via a Newton-Raphson
iterative algorithm for velocities in the range $0.1 \leq v < 0.99$.
Note that $v=1$ is the familiar magnon velocity (in rationalized units)
and provides an upper bound for the existence of a solitary wave
in rigid motion. On the contrary, there is no lower bound for the velocity -- the
restriction $v \geq 0.1$ was dictated only by numerical expedience.
Now, the calculated solitary wave is illustrated in Fig.~\ref{fig:kelvin02}
for the relatively low velocity $v=0.2$ and does indeed describe a VA pair
(with $\pontryagin = 0$) at a relative distance $d=4.97\sim 1/v$,
as anticipated by the discussion of model VA pairs in Section~\ref{sec:model}
and the heuristic asymptotic analysis earlier in this section. But we are now in
a position to carry out an accurate calculation practically throughout
the allowed range $0< v < 1$, to discover that there exists a characteristic velocity
$v_0 \sim 0.78$ above which the vortex-antivortex character is lost ($d\sim 0$)
and the solitary wave becomes a lump with no apparent topological features,
as illustrated in Fig.~\ref{fig:kelvin095} for $v=0.95$.

\begin{figure}
\epsfig{file=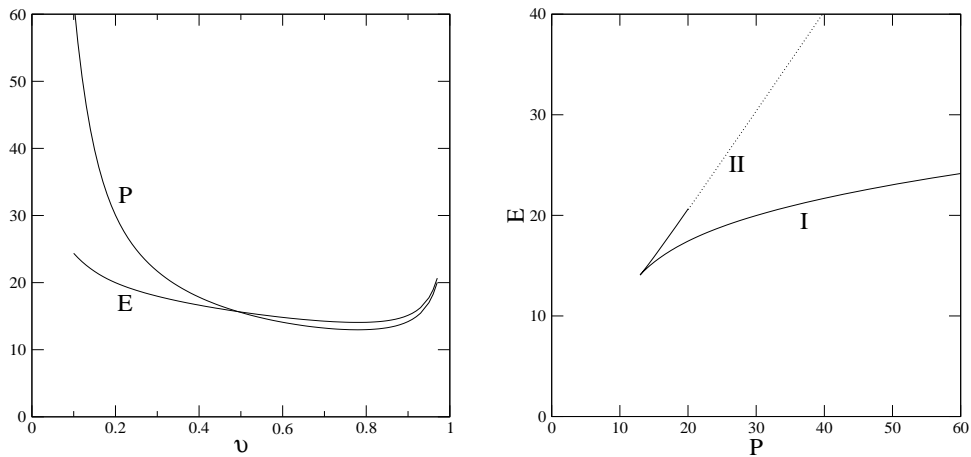,width=0.80\linewidth}
   \caption{Energy $E$ and impulse $P$ as functions of velocity $v$
(left panel) and $E$ vs $P$ dispersion (right panel)
for a solitary wave in Kelvin motion.
The dotted line was calculated from the asymptotic dispersion
(\ref{eq:kelvinpairenergy2}) applied for $Q_0=4.7$.
   }  \label{fig:kelvin_energy}
\end{figure}

The existence of a characteristic velocity $v_0$ becomes
apparent when we calculate energy $E$ and impulse $P$ as functions of
$v$, as shown in Fig.~\ref{fig:kelvin_energy}.
Note that both $E$ and $P$ develop a minimum at a common velocity $v=v_0=0.78$.
As a result, the energy vs impulse dispersion shown in
Fig.~\ref{fig:kelvin_energy}
develops a cusp at a point $(E_0,P_0)$ that corresponds to the
values of $E$ and $P$ at $v=v_0$.
Thus the calculated family of solitary waves consists of two branches.
Branch I consists of VA pairs that propagate with
velocities in the range  $v < v_0$.
The corresponding branch in the dispersion of Fig.~\ref{fig:kelvin_energy}
approaches the asymptotic dispersion (\ref{eq:kelvinpairenergy})
for large $P$ (or $v\to 0$). Indeed, an excellent fit of the data
is obtained by Eq.~(\ref{eq:kelvinpairenergy}) if we choose the subleading
constant according to $\ln P_0 \sim 1/4$.
Branch II consists of lumps with no apparent topological features
which propagate with velocities in the range $v_0 < v < 1$.
The corresponding branch in the dispersion of Fig.~\ref{fig:kelvin_energy}
is accurately described in the asymptotic ($v\to 1$) region, where
$P$ again becomes large, by
\begin{equation}  \label{eq:kelvinpairenergy2}
E = P\, \left(1 + \frac{Q_0^2}{2P^2} + \ldots\right)
\end{equation}
with $Q_0\approx 4.7$. Actually this asymptotic dispersion can be derived by
showing that in the limit $v \to 1$ the 2D Landau-Lifshitz equation
reduces to what is a modified Kadomtsev-Petviashvili (KP)
equation \cite{papanicolaou99}.

A cusp in the energy vs impulse dispersion
occurred previously in a calculation of vortex rings
in a model superfluid by Jones and Roberts \cite{jones82}.
The same authors together with Putterman later argued that the solitary waves
that correspond to branch II are actually  unstable \cite{jones86}.
Interestingly, a dispersion with a cusp appears also in the much simpler
problem of electron-positron motion in a uniform magnetic field,
where the motion that corresponds to branch II is also unstable
(see our Appendix).
Hence, while a stability analysis has not yet been carried out for the solitary
waves described in this section, it is reasonable to expect that the lumps
of branch II may be unstable.
But there is every reason to believe that the Kelvin motion of the VA pairs
of branch I is indeed stable.

\section{Rotational motion}
\label{sec:rotate}

The possibility of topologically nontrivial ($\pontryagin=\pm 1$) VA pairs
in steady rotational motion was recently examined by one of us \cite{komineas07}.
Again, one may invoke the model VA pair of Section~\ref{sec:model}
to understand some important features of the rotational motion
at large distance $d$.
For definiteness we consider a VA pair defined by Eq.~(\ref{eq:productomega})
with $\kappa_1=-\kappa_2=1$ and $\lambda_1=-\lambda_2=-1$, thus $\pontryagin = 1$.
The local topological vorticity $\gamma$ is then peaked around the positions of the
vortex and the antivortex, now with weight equal to $2\pi$ in both cases.
Then, for large $d$, the angular momentum defined by Eq.~(\ref{eq:angmomentum})
is estimated to be $\am \sim \half 2(2\pi)(\frac{d}{2})^2=\frac{\pi}{2} d^2$.
One may also consider the Derrick-like scaling relation applied to the extended
energy functional $F=E-\omega \am$:
\begin{equation}  \label{eq:derrick2}
 \omega \am = \half\int{m_3^2\, dx dy} = \Ea
\end{equation}
where $\Ea$ is the total anisotropy energy of a VA pair which approaches
asymptotically $\Ea \sim (\frac{\pi}{2}+\frac{\pi}{2}) = \pi$, and hence $\omega \am = \pi$,
because the anisotropy energy of a single vortex is equal to  $\pi/2$
according to Eq.~(\ref{eq:virial1}).
To summarize,
\begin{equation}  \label{eq:estimates2}
\am \sim \frac{\pi}{2}\, d^2\,,\quad \omega \am \sim \pi\,,\quad \omega \sim \frac{2}{d^2}.
\end{equation}
One may further employ the familiar relation
\begin{equation}  \label{eq:dedl}
\omega = \frac{dE}{d\am}
\end{equation}
in which we insert the estimate $\omega\sim\pi/\am$ to obtain an elementary differential
equation for $E$ whose integral is
\begin{equation}  \label{eq:rotatepairenergy}
E \approx \pi\ln(\am/\am_0)
\end{equation}
where $\am_0$ is an integration constant that cannot be fixed by the present
leading-order argument. Nevertheless, Eq.~(\ref{eq:rotatepairenergy})
provides the essence of the energy vs angular momentum
dispersion for large relative distance $d$ or small angular frequency
$\omega \sim 2/d^2$ and hence large angular momentum $\am\sim\frac{\pi}{2}\,d^2$.

While the preceding heuristic asymptotic analysis is very useful for
understanding some basic aspects of a rotating VA pair, it does not give us any clue
concerning  the fate of the pair at small vortex-antivortex separation.
In principle, such a question could be settled by solving numerically
the analog of Eq.~(\ref{eq:lleomega1}) for a solitary wave rotating
at constant angular frequency $\omega$:
\begin{equation}  \label{eq:llerotateomega}
i\omega\,\eab\,x_\alpha\p_\beta\Omega + \Delta\Omega
  + \frac{1-\bOmega\Omega}{1+\bOmega\Omega}\,\Omega =
\frac{2\bOmega}{1+\bOmega\Omega}\,(\bnabla\Omega\cdot\bnabla\Omega).
\end{equation}
A numerical solution could again be attempted by an iterative Newton-Raphson
algorithm. Actually, in this case, we found it more convenient to employ
a relaxation algorithm to derive approximate numerical solutions as stationary
points of the extended energy functional $F=E-\omega \am$ \cite{komineas07}.

\begin{figure}
\epsfig{file=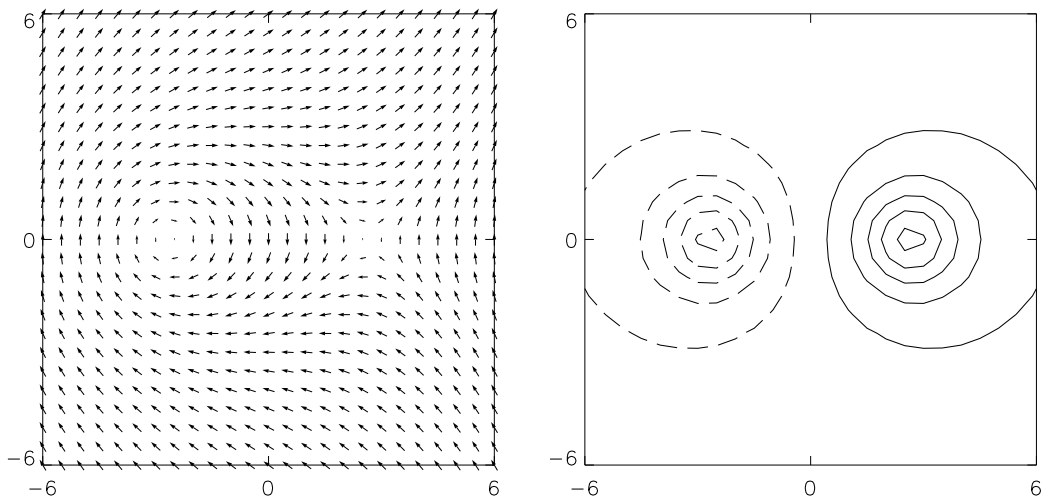,width=0.9\linewidth}
  \caption{Snapshot of a topologically nontrivial ($\pontryagin=1$)
VA pair in rotational motion (conventions as in Fig.~\ref{fig:kelvin02},
with solid lines in the right panel corresponding to positive values
of $m_3$ and dashed lines to negative ones).
The pair rotates around a fixed guiding center taken at the origin
of coordinates, with angular velocity $\omega=0.06$ and calculated
relative distance $d=5.3$, energy $E=21$, and angular impulse $\am=64$.
}  \label{fig:rotate700}
\end{figure}

\begin{figure}
\epsfig{file=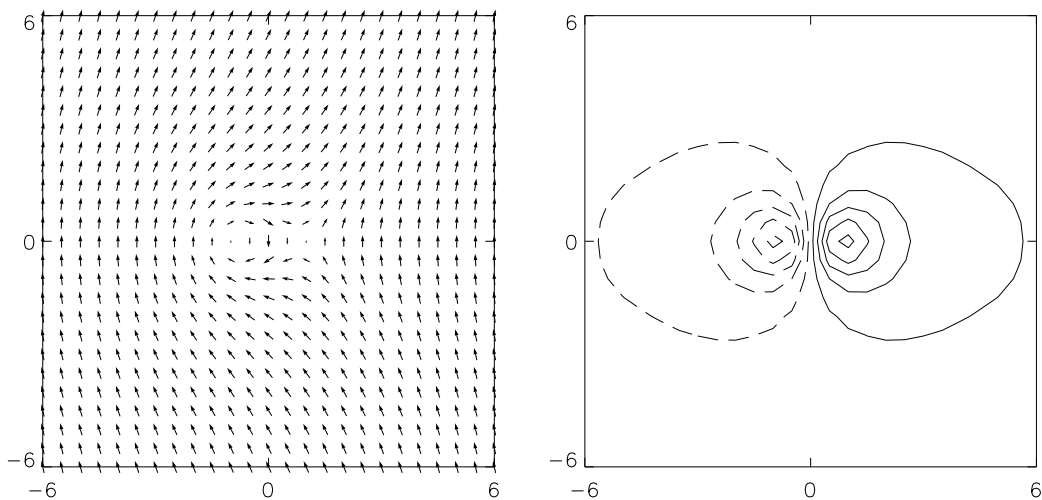,width=0.9\linewidth}
  \caption{$\pontryagin=1$ VA pair in rotational motion with
angular velocity $\omega=0.18$ and calculated
relative distance $d=1.7$, energy $E=15$, and angular impulse $\am=11$.
The only difference from Fig.~\ref{fig:rotate700} is that the
overall size of the pair is now reduced.
}  \label{fig:rotate200}
\end{figure}

The calculated configuration for $\omega=0.06$ is shown in Fig.~\ref{fig:rotate700}
and does indeed correspond to a topologically nontrivial ($\pontryagin=1$)
rotating VA pair consisting of a vortex with negative polarity $(\kappa,\lambda)=(1,-1)$
and an antivortex with positive polarity $(\kappa,\lambda)=(-1,1)$,
as anticipated by the general discussion of Section~\ref{sec:model}.
In particular, for this relatively small value of $\omega$,
the calculated distance $d$ in the rotating pair is relatively large
$(d=5.3)$, so is the angular momentum $(\am=64)$;
in rough agreement with the asymptotic estimates of Eq.~(\ref{eq:estimates2}).

\begin{figure}
\epsfig{file=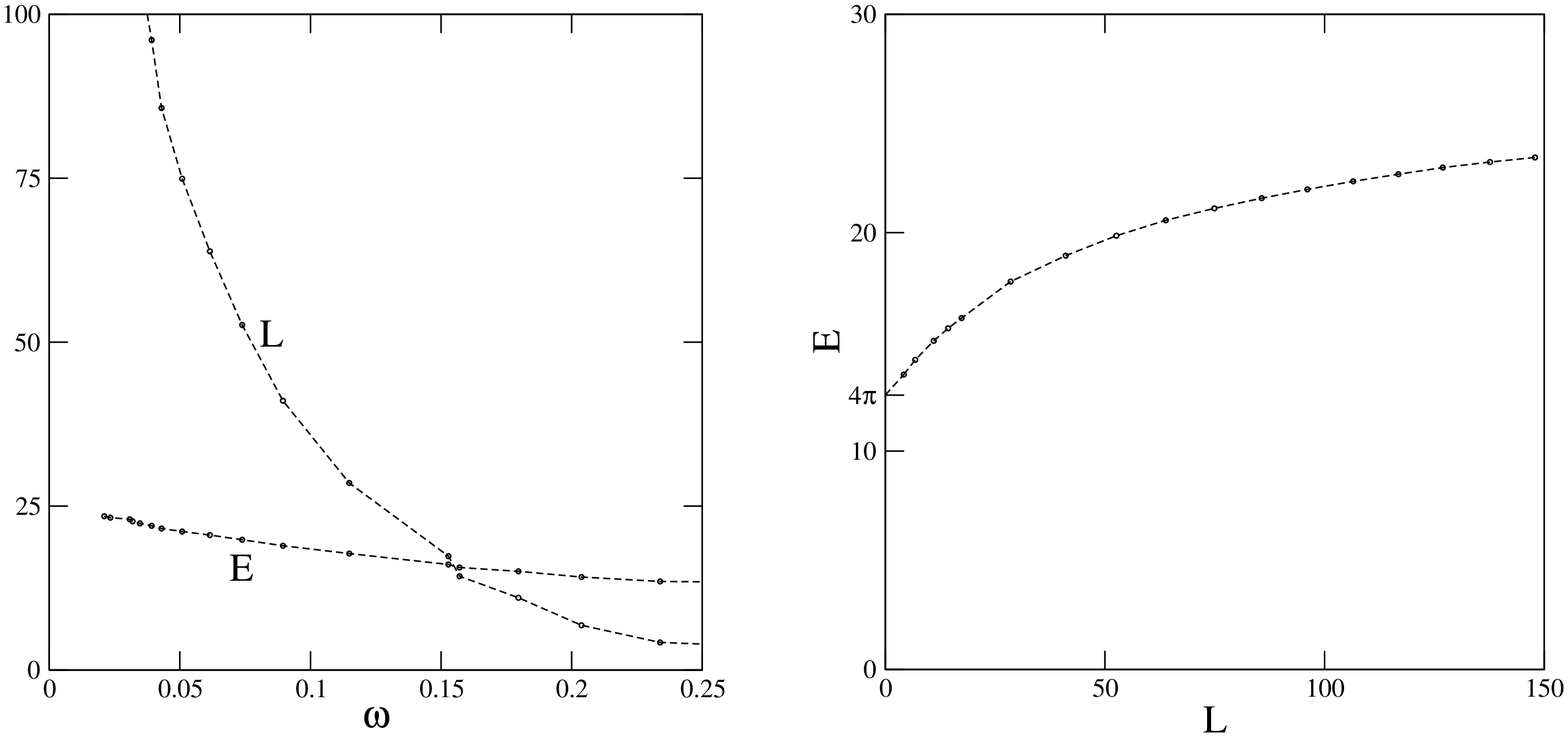,width=0.80\linewidth}
   \caption{Energy $E$ and angular impulse $\am$ as functions of
angular velocity $\omega$ (left panel) and $E$ vs $\am$ dispersion (right panel)
for an $\pontryagin=1$ VA pair in rotational motion.
   }  \label{fig:rotate_energy}
\end{figure}

The calculation was repeated for a larger value of angular velocity
$(\omega=0.18)$ to find that both the relative distance $(d=1.7)$
and the angular momentum $(\am=11)$ are reduced to smaller values.
But the general structure of the solution shown in Fig.~\ref{fig:rotate200}
for $\omega=0.18$ remains basically the same as that for $\omega=0.06$,
except that the overall size of the VA pair is reduced.
For yet larger values of $\omega$ the angular momentum $\am$ tends to vanish.
This trend is apparent in Fig.~\ref{fig:rotate_energy} which illustrates the
dependence of $E$ and angular momentum $\am$ as functions of angular velocity $\omega$,
as well as the $E$ vs $\am$ dispersion. For large values of $\am$ the above
dispersion exhibits logarithmic dependence, as predicted by the asymptotic
result of Eq.~(\ref{eq:rotatepairenergy}). But the most important feature
of the calculated dispersion is that energy approaches the finite value
$E=4\pi$ as $\am\to 0$, while the corresponding rotating VA pair becomes vanishingly
small. This result is of central importance for our main argument and will
be analyzed in some detail in the remainder of this section.

The best way to describe a vanishing VA pair, in the limit $\am\to 0$,
is to invoke yet another model configuration through the stereographic
variable
\begin{equation}  \label{eq:VAskyrmion}
\Omega = \frac{\bzeta-\frac{d}{2}}{\bzeta+\frac{d}{2}}\,,\quad
\zeta = x+i y\,,\;\; \bzeta = x - i y\,,
\end{equation}
where the constant $d$ is taken to be real for simplicity.
Now, configuration (\ref{eq:VAskyrmion}) reaches a finite value $\Omega=1$
at spatial infinity and the corresponding magnetization is uniform:
\begin{equation}  \label{eq:bc}
\bm{m} = (1,0,0)\quad as \quad |\zeta| \to \infty
\end{equation}
which is an appropriate boundary value (modulo a constant azimuthal rotation)
for easy-plane anisotropy discussed here.
Because of (\ref{eq:bc}) the skyrmion number is expected to be an integer
and is actually computed to be $\pontryagin=1$ by a direct application
of Eq.~(\ref{eq:skyrmionnumber}). Furthermore, the spin configuration derived from
(\ref{eq:VAskyrmion}) is an exact solution of the LL equation, if we neglect
anisotropy, with exchange energy
\begin{equation}  \label{eq:skyrmionenergy}
E=\Ee = \half \int \left(\nabla\bm{m}\cdot\nabla\bm{m}\right)\, dx dy = 4\pi\,,
\end{equation}
for any $d$, as discussed long time ago by Belavin and Polyakov \cite{belavin75}.

We now return to the main line of argument by noting that configuration
(\ref{eq:VAskyrmion}) may also be thought of as a topologically nontrivial
($\pontryagin=1$) model VA pair consisting of a 
$(\kappa,\lambda)=(1,-1)$ vortex and a $(\kappa,\lambda)=(-1,1)$ antivortex
located at a distance $d$ apart, in close analogy with the model VA pair
constructed in Section~\ref{sec:model}.
Although (\ref{eq:VAskyrmion}) is not an exact solution in the presence of
anisotropy, it provides a good model for the behavior of a vanishing VA pair
in the limit $\am\to 0$. Anisotropy sets a distance scale $R\sim 1/\sqrt{\qfactor} = 1$
(in rationalized units) beyond which a physically acceptable configuration
must reach the uniform value (\ref{eq:bc}).
In a sense, this property is shared by configuration (\ref{eq:VAskyrmion})
because it becomes uniform almost everywhere when $d \ll 1$ while it retains
its topological structure as long as $d$ remains finite.
The strict limit $d\to 0$ is not uniform.
If taken naively, all topological structure appears to be lost and both
energy $E$ and skyrmion number $\pontryagin$ appear to vanish.
However, if integrals are performed before taking the $d\to 0$ limit,
$E$ approaches $4\pi$ without encountering an energy barrier
while $\pontryagin=1$ for all $d$.
Clearly the limit $d\to 0$ creates a singular point which hides all
topological structure. The main point is the claim that a similar
situation arises in the calculated rotating VA pair in the limit $\am\to 0$,
as discussed further in Section~\ref{sec:switching}.

This section is completed with a comment concerning the manner in
which the $\am\to 0$ limit is reached. Our numerical data as well as
virial relation (\ref{eq:derrick2}) are consistent with a linear dispersion
\begin{equation}  \label{eq:rotatedispersion}
E \approx 4\pi + \half\,\am
\end{equation}
in the limit $\am\to 0$, which implies a finite value of the angular velocity
$\omega = dE/d\am=1/2$ in the limit of a vanishing VA pair,
and thus an upper limit $\omega_{\rm max} \approx  1/2$.

\section{Vortex core switching}
\label{sec:switching}

Applied for long time intervals, our relaxation algorithm
revealed some tendency for instability of a rotating VA pair,
probably due to radiation effects analogous to those expected
for a pair of rotating electric charges discussed in the Appendix.
However, the basic features of rotating VA pairs studied in the
preceding section persist over sufficiently long time intervals
and are thus relevant for practical applications.
In fact, in a realistic ferromagnet, some dissipation is always
present and can be modeled by introducing Gilbert damping
in the LL equation through the replacement
\begin{equation}  \label{eq:gilbert}
\frac{\p \bm{m}}{\p t} \to \frac{\p \bm{m}}{\p t} +
 \alpha \left(\bm{m}\times \frac{\p \bm{m}}{\p t}\right)
\quad \rm{or}\quad
i\,\frac{\p \Omega}{\p t} \to (i-\alpha)\,\frac{\p \Omega}{\p t}
\end{equation}
in Eq.~(\ref{eq:lle}) or Eq.~(\ref{eq:lleomega}), respectively,
where $\alpha$ is a dissipation constant.

The dynamics of a topologically nontrivial ($\pontryagin=\pm 1$)
VA pair may thus be summarized as follows. The vortex and the antivortex rotate
around each other, while the pair  shrinks due to dissipation.
The energy of the pair follows approximately the curve
of the right panel of Fig.~\ref{fig:rotate_energy} as its size (and its angular momentum)
decreases. At vanishing size a singular point of the type discussed
in the preceding section would be created and the total energy
would reach the finite value $E=4\pi$ (in rationalized units).
However, the discreteness of the lattice actually interrupts the process
when the size of the pair becomes comparable to the lattice spacing.
The VA pair disappears (i.e., the skyrmion number changes
abruptly from $\pontryagin=\pm 1$ to $\pontryagin=0$) and a burst
of energy equal to $4\pi$ is released into the system,
probably in the form of spin waves.
In physical units, the amount of energy released is given by $E=8\pi t A$
where $t$ is the film thickness and $A$ the exchange constant.
For typical values $t=10$ nm and $A=10^{-11}$ J/m we obtain the estimate
$E \sim 2.5\times 10^{-18}$ J which is apparently in rough
agreement with numerical simulations \cite{hertel06,tretiakov07}.

The scenario described above explains how a topologically forbidden
$(\Delta\pontryagin = 1)$ transition can take place in a real ferromagnet
but does not by itself account for the experimentally observed
vortex core switching. The complete scenario involves two distinct steps.
First, application of a short burst of an alternating magnetic field
creates a VA pair in the vicinity of a preexisting single vortex.
Second, a three-body collision takes place during which a suitable
pair of vortices is annihilated through a $\Delta\pontryagin = 1$
transition of the type described above, and the final product is a
single vortex with polarity opposite to that of the original vortex
\cite{waeyenberge06}.
We also note that a system with three vortices
(two vortices and an antivortex which form a cross-tie wall)
was observed in a rectangular platelet in \cite{kuepper07}.
Their spectrum of eigenmodes was studied experimantally.

   Here we do not address the question of how a VA pair is actually
created. Rather we concentrate on step 2 of the process and explain in some
detail how vortex core switching may occur in a three-body collision.
Specifically, let us assume that a single $(1,-1)$=C vortex
is initially at
rest at some specified point which is taken to be the origin of the
coordinate system. Let us further assume that a topologically trivial $(\pontryagin =0)$
VA pair consisting of a $(1,1)$=A vortex and a $(-1,1)$=B antivortex is somehow
created in the neighborhood of the original vortex. Once created the AB pair
will undergo Kelvin motion of the type described in Section \ref{sec:kelvin} and eventually
collide with the single vortex C.

\begin{figure}
\epsfig{file=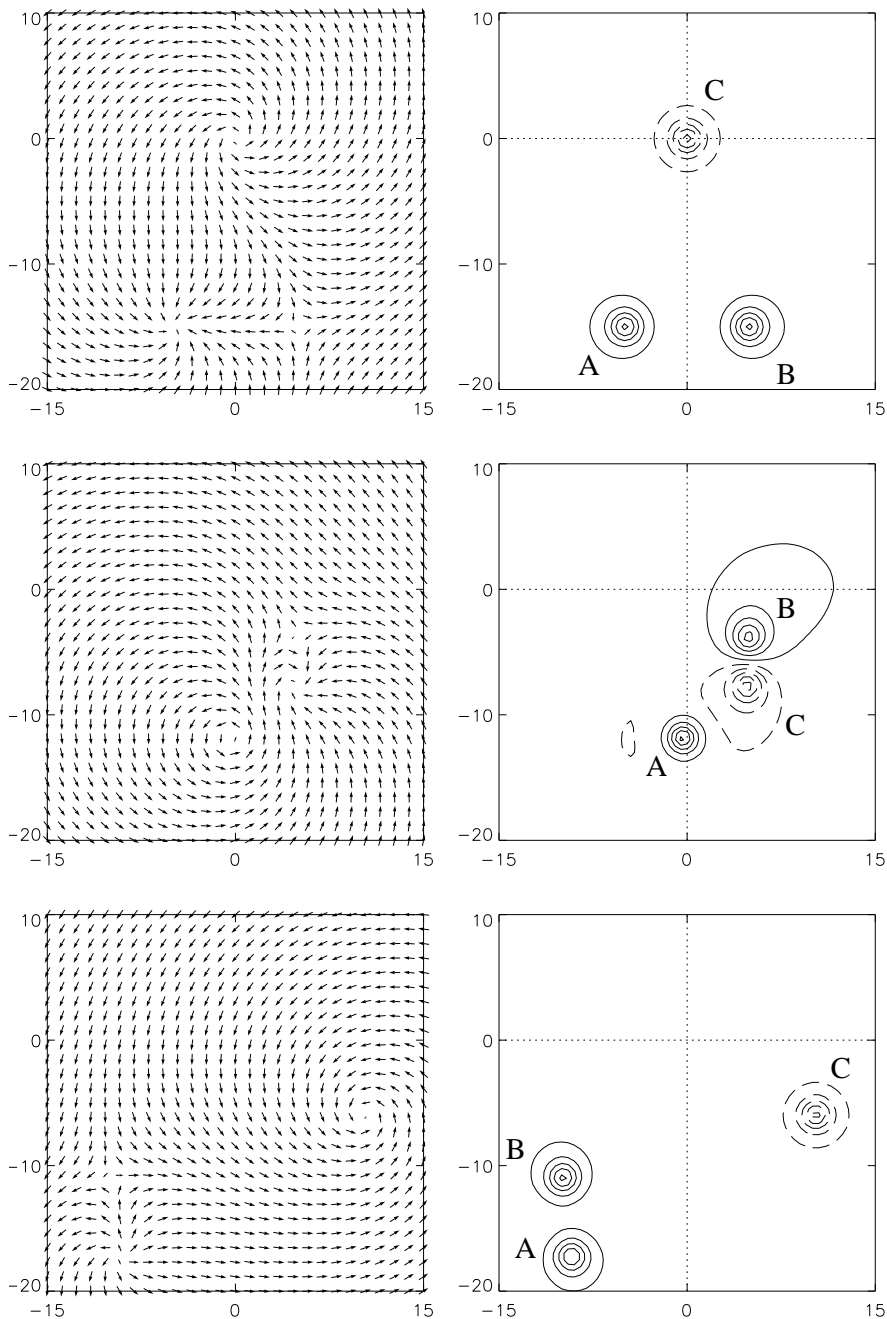,width=0.75\linewidth}
   \caption{Three snapshots for the collision of a VA pair in Kelvin motion
(the AB pair), initially located at $(0,-15)$ and propagating
with velocity $v=0.1$, against a target vortex C initially located
at the origin.
%at times t=0, 70, and 170.
During collision, antivortex B rotates around vortex C before rejoining
its original partner A to form a new VA pair that scatters off at an angle
in the third quadrant.
The target vortex C is shifted to a
new position in the fourth quadrant thanks to transmutation
of VA pair momentum to vortex position.
   }  \label{fig:switch_v01}
\end{figure}

   The process was simulated by a numerical solution of the
corresponding initial-value problem in the LL equation. Figure \ref{fig:switch_v01}
provides an illustration with three characteristic snapshots in the case of a relatively
slow Kelvin pair initially moving along the y axis with velocity $v=0.1$ for
which the vortex and the antivortex are separated by a distance
$d \approx 1/v=10$. As the pair approaches, the original C=$(1,-1)$ vortex teams
up with the B=$(-1,1)$ partner of the AB pair to form a new, topologically
nontrivial $(\pontryagin =1)$ VA pair in quasi-rotational motion. In fact, B rotates
almost a full circle around C before rejoining its original partner A. The new
AB pair is again a topologically trivial $(\pontryagin =0)$ VA pair in Kelvin motion that
moves away from the target vortex, having suffered a total scattering angle
that is greater than $\pi/2$ from its original direction.
The scattering is
inelastic in the sense that the outgoing AB pair moves out with greater
velocity $(v=0.15)$. And, most remarkably, the target vortex C moves away from
the origin and comes to rest at a new location in the fourth quadrant of the
xy plane.

   This rather unusual behavior is explained by the unusual nature of
the conservation laws (11) and (12) which allow for a transmutation between
position and impulse in the case of topologically nontrivial systems, such
as the three-vortex system considered here (a more detailed discussion
will be given in a future publication).

\begin{figure}
\epsfig{file=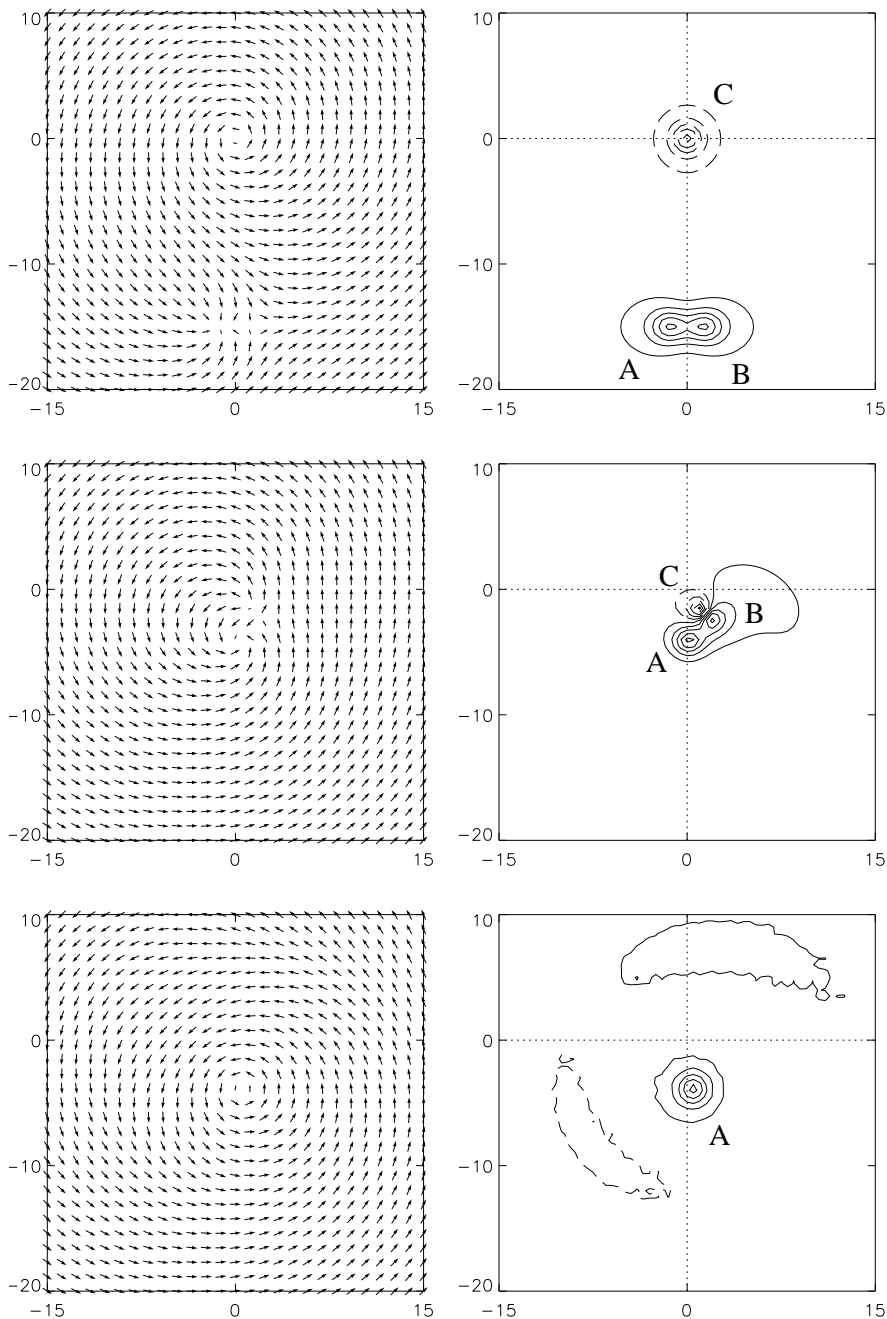,width=0.75\linewidth}
   \caption{
Same as Fig.~\ref{fig:switch_v01} but for a larger initial velocity $v=0.5$ of
the AB pair. During collision, antivortex B begins
to rotate around vortex C but the rotating BC pair is
eventually annihilated leaving behind vortex A (with
polarity opposite to that of the target vortex C) and
a burst of spin waves that propagate away from the scattering region.
%times t=0, 24, and 32.
   }  \label{fig:switch_v05}
\end{figure}

   The preceding numerical experiment was repeated for a Kelvin pair
with relatively large velocity $v=0.5$ for which the vortex and the antivortex
are tightly bound at a relative distance $d=2.6$ \cite{papanicolaou99}. The process is
again illustrated by three characteristic snapshots in Figure \ref{fig:switch_v05}.
While the initial stages of the process are similar to those encountered in the case
of slow Kelvin motion (Figure \ref{fig:switch_v01}) a substantial departure occurs when the
pair now approaches the target vortex. In particular, as soon as antivortex
B=$(-1,1)$ begins to rotate around the target vortex C=$(1,-1)$ they collide and
undergo a spectacular $\Delta\pontryagin =1$ transition (annihilation) leaving behind the
A=$(1,1)$ vortex which may be thought of as the target vortex C=$(1,-1)$ with
polarity flipped from $-1$ to 1 (vortex core switching) and a burst of spin
waves propagating away from the scattering region.

  A detailed numerical investigation of the three-vortex process for
Kelvin waves with velocities in the allowed range $0<v<1$ suggests the
existence of the three characteristic regions separated by two critical
velocities $v_1=0.3$ and $v_2=0.9$ (such that $0<v_1<v_0<v_2<1$, with $v_0=0.78$ being the
critical velocity discussed in Section \ref{sec:kelvin}). For $0<v<v_1$, the Kelvin pair
undergoes nearly elastic scattering of the type depicted in Figure \ref{fig:switch_v01}.
For $v_1<v<v_2$, the process leads to a topologically forbidden $\Delta\pontryagin =1$
transition of the type illustrated in Figure \ref{fig:switch_v05}.
There is also some evidence that fast
Kelvin waves with velocities in the narrow range $v_2<v<1$ undergo a nearly
elastic scattering without inversion of the polarity of the target vortex.

\section{Conclusion}
\label{sec:conclusion}

The VA pairs analyzed in this paper are special examples of solitary waves
whose dynamics is closely related to their topological structure.
For example, the VA pairs studied in Section~\ref{sec:kelvin}
can undergo free translational motion because their topological charge
vanishes ($\pontryagin = 0$).
In contrast, a VA pair with nonvanishing $\pontryagin$ performs
rotational motion around a fixed guiding center and is thus spontaneously
pinned within the ferromagnetic medium, as discussed
in Section~\ref{sec:rotate}.
Such a peculiar dynamical behavior would have been surprising had it not
occurred previously in the case of interacting electric charges in the
presence of a magnetic field.

It is then interesting to ascertain conditions under which
a certain field theory would exhibit a similar link between
topology and dynamics. A simple criterion was introduced in Ref.~\cite{komineas98}
and is briefly summarized as follows.
We restrict attention to 2D Hamiltonian systems described in terms of
$\Lambda$ pairs of canonically conjugate variables $(\Pi_i,\Phi_i)$
with $i=1,2,\ldots \Lambda$. Then one may define the local vorticity
\begin{equation}  \label{eq:vorticity2}
\gamma = \sum_{i=1}^\Lambda \eab\, \p_\alpha\Pi_i\, \p_\beta\Phi_i
\end{equation}
which is a simple generalization of the first step of Eq.~(\ref{eq:vorticity1}),
the remaining two steps being special to the specific example
considered in the present paper (ferromagnets).
Now, in general, the total vorticity $\Gamma=\int \gamma\, dx dy$ is expected
to vanish by a trivial partial integration using Eq.~(\ref{eq:vorticity2}).
On the other hand, a nonzero $\Gamma$ would signal a special topological structure
of the field theory under consideration and may lead to peculiar dynamics.
The theory analyzed in the present paper is an example (with $\Lambda=1$) which yields
a nonzero $\Gamma$ that may be identified with the Pontryagin index
$\pontryagin = \Gamma/4\pi$. An example with $\Lambda=2$ is provided by
a 2D antiferromagnet where $\Gamma = 0$ except when an external field
is present which may lead to $\Gamma \neq 0$ and an interesting link
between topology and dynamics \cite{komineas98}.

Implicit in the preceding general argument is the fact that a topological charge
$\pontryagin$ is conserved. Also taking into account that $\pontryagin$
is quantized, one would expect that a topological ($\pontryagin \neq 0$) soliton
cannot be annihilated in a continuous manner.
Nevertheless, a quasi-continuous process was described in Ref.~\cite{komineas07}
and in the present paper according to which a rotating VA pair with $\pontryagin = 1$
may be reduced to a singular point and thereby be eliminated by lattice
discreteness without encountering an energy barrier.

A mechanism for changing the topological number of a magnetic
configuration makes it possible to obtain controlled switching
between topologically distinct (and thus robust) magnetic states.
This was achieved in the experiments of Refs.~\cite{waeyenberge06,yamada07}.
The dynamics underlying both experiments involves a three-vortex process
\cite{hertel07} initiated by the production of a topologically trivial
VA pair in the vicinity of a preexisting vortex by an
alternating magnetic field.
It has been predicted that the same phenomenon would occur
if one uses a rotating external field \cite{kravchuk07}.
The resulting three-vortex system carries nonzero topological charge
and is thus by itself a rotating object spontaneously pinned
in the magnet.
Also due to dissipation, a quasi-continuous process takes place
that changes the topological number by one unit,
leaving behind a burst of energy in the form of spin waves
and a single vortex with polarity opposite to that of the original vortex.

Although our strictly 2D treatment provides a detailed scenario for the
three-vortex process that leads to polarity switching, it does not
account for the initial production of a topologically trivial VA pair.
This is partly due to our approximation of a thin film with infinite extent.
Implicit in this approximation is the assumption that the demagnetizing
field amounts to a simple (additive) renormalization of easy-plane
anisotropy. While this assumption appears to be firmly established
for static magnetic states \cite{gioia97} we do not know of a
corresponding mathematical derivation for dynamical processes of the
type discussed here.

It is also important to visualize how the formation of a singular
point discussed in the present strictly 2D context appears within
a realistic magnetic element of finite extent.
Numerical simulations \cite{hertel07} show that, in an element
of finite thickness, a singular point is first created at one of the
surfaces of the element. The VA pair then vanishes by formation
and subsequent annihilation of a singular point at successive
levels away from the surface. At the stage when a singular point has
been formed and annihilated, say, near the top surface, while the
VA pair is still present in the bulk of the element,
a Bloch Point (BP) is created in the element.
This is a somewhat simplified realization of the BP studied
in Ref.~\cite{doering68}.
Needless to say, the BP created near the top surface is eventually
annihilated when the VA pair exits the system through the lower surface.
It is important to emphasize the unusual fact that during
creation and annihilation of the BP the system does not have to overcome
an energy barrier, unlike the case discussed in \cite{slonczewski75},
essentially for the same reasons explained for the strictly
2D VA pairs studied in the main text.

\begin{acknowledgments}
N.P. is grateful for hospitality at the Max-Planck Institute
for the Physics of Complex Systems (Dresden) where this work was completed.
\end{acknowledgments}

\begin{appendix}
\section{Electric charges in a magnetic field}

Most of the distinct features of the dynamics of VA pairs occur also
in the dynamics of electric charges in the presence of a uniform magnetic field $\bm{B}$.
Although we shall mainly be interested in 2D motion in a plane perpendicular to $\bm{B}$,
it is convenient to keep for the moment 3D notation and write the equations
of motion for two interacting charges $e_1$ and $e_2$:
\begin{equation}  \label{eq:A1}
m\, \frac{d\bm{v}_1}{d t} = \bm{F}_1 + e_1 (\bm{v}_1\times\bm{B})\,,\quad
m\, \frac{d\bm{v}_2}{d t} = \bm{F}_2 + e_2 (\bm{v}_2\times\bm{B})\,,
\end{equation}
where
\begin{equation}  \label{eq:A2}
\bm{F}_1 = -\bm{F}_2 = -\frac{\bm{r}_1-\bm{r}_2}{|\bm{r}_1-\bm{r}_2|}\;
                                          V'(|\bm{r}_1-\bm{r}_2|)
\end{equation}
is the mutual force derived from a potential energy $V=V(|\bm{r}_1-\bm{r}_2|)$
and $V'$ denotes derivative with respect to the argument $|\bm{r}_1-\bm{r}_2|$.
The conserved energy functional is then given by
\begin{equation}  \label{eq:A3}
E = \half m (v_1^2 + v_2^2) + V(|\bm{r}_1-\bm{r}_2|),
\end{equation}
which does not depend explicitly on the magnetic field, while the conserved
linear momentum (impulse) is now given by
\begin{equation}  \label{eq:A4}
\bm{P} = m (\bm{v}_1+\bm{v}_2) - (e_1\bm{r}_1 + e_2 \bm{r}_2)\times \bm{B}
\end{equation}
and differs from the usual mechanical definition by an important field dependent term.
This (second) term actually indicates a rather profound influence of the magnetic field
on the dynamics of electric charges.
For instance, note that a shift of the origin of coordinates by a constant vector
$\bm{c}$, thus $\bm{r}_1 \to \bm{r}_1 - \bm{c}$
and $\bm{r}_2 \to \bm{r}_2 - \bm{c}$, induces a nontrivial change on the impulse
$\bm{P}$ of Eq.~(\ref{eq:A4}) given by
\begin{equation}  \label{eq:A5}
\bm{P} \to \bm{P} + (e_1+e_2) (\bm{c}\times \bm{B}).
\end{equation}
This unusual behavior is analogous to that of the impulse defined
by Eq.~(\ref{eq:linmomentum}) in the case of field configurations
with nonvanishing total topological vorticity $\Gamma$ or skyrmion number $\pontryagin$.
Here we may abstract from Eq.~(\ref{eq:A5}) the analog of the topological vorticity
in the present problem:
\begin{equation}  \label{eq:A6}
\Gamma \sim (e_1+e_2)\,B.
\end{equation}
An electron-positron pair ($\Gamma=0$) may undergo Kelvin motion,
while two like charges ($\Gamma \neq 0$) perform rotational motion around a fixed guiding
center, as determined by the explicit solutions
constructed and briefly analyzed in the remainder of this appendix.

Consider first the case of an electron-positron pair ($e_1=-e_2=e$) for which
a special 2D solution of Eqs.~(\ref{eq:A1}) is given by
\begin{eqnarray}  \label{eq:A6a}
 x_1 = \frac{d}{2}\,,\quad y_1 & = & v t\,,\quad z_1=0\,,  \nonumber \\
 x_2 = -\frac{d}{2}\,,\quad y_2 & = & v t\,,\quad z_2=0\,,
\end{eqnarray}
where the electron and the positron are located at a constant
relative distance $d$ along the x axis and move in formation
along the y axis with constant velocity
\begin{equation}  \label{eq:A7}
 v = \frac{V'(d)}{e\,B}\,,
\end{equation}
in close analogy with the Kelvin motion of the VA pair discussed
in Section~\ref{sec:kelvin}.
We further calculate energy $E$ from Eq.~(\ref{eq:A3}) and impulse
$\bm{P}=(0,P,0)$ from Eq.~(\ref{eq:A4}) to find
\begin{eqnarray}  \label{eq:A8}
 E & = & m v^2 + V(d) = m \left[\frac{V'(d)}{eB}\right]^2 + V(d) \nonumber \\
 P & = & 2 m v + eBd  = 2m \frac{V'(d)}{eB} + eBd.
\end{eqnarray}
Hence all relevant quantities are given in parametric form as functions
of relative distance $d$ once the potential energy $V=V(d)$ has been specified.
But there are some generic properties of this solution that are practically
independent of the choice of $V(d)$.
We may take derivatives with respect to $d$ of both sides
of Eq.~(\ref{eq:A8}) to find $E' = V'(1+2 m V''/e^2B^2)$ and
$P' = eB(1+2 m V''/e^2B^2)$.
An immediate consequence of these relations is the group-velocity
relation $v=dE/dP$. We also note that both $E$ and $P$ may acquire an extremum
at a common value of $d$ (or $v$) determined from
\begin{equation}  \label{eq:A9}
 1 + \frac{2m V''(d)}{e^2B^2} = 0.
\end{equation}
This is actually the reason for the appearance of a cusp in the $E$ vs $P$ dispersion
analogous to that encountered in the Kelvin motion of VA pairs, which now appears
to be a generic feature of a wide class of physical systems.

\begin{figure}
\epsfig{file=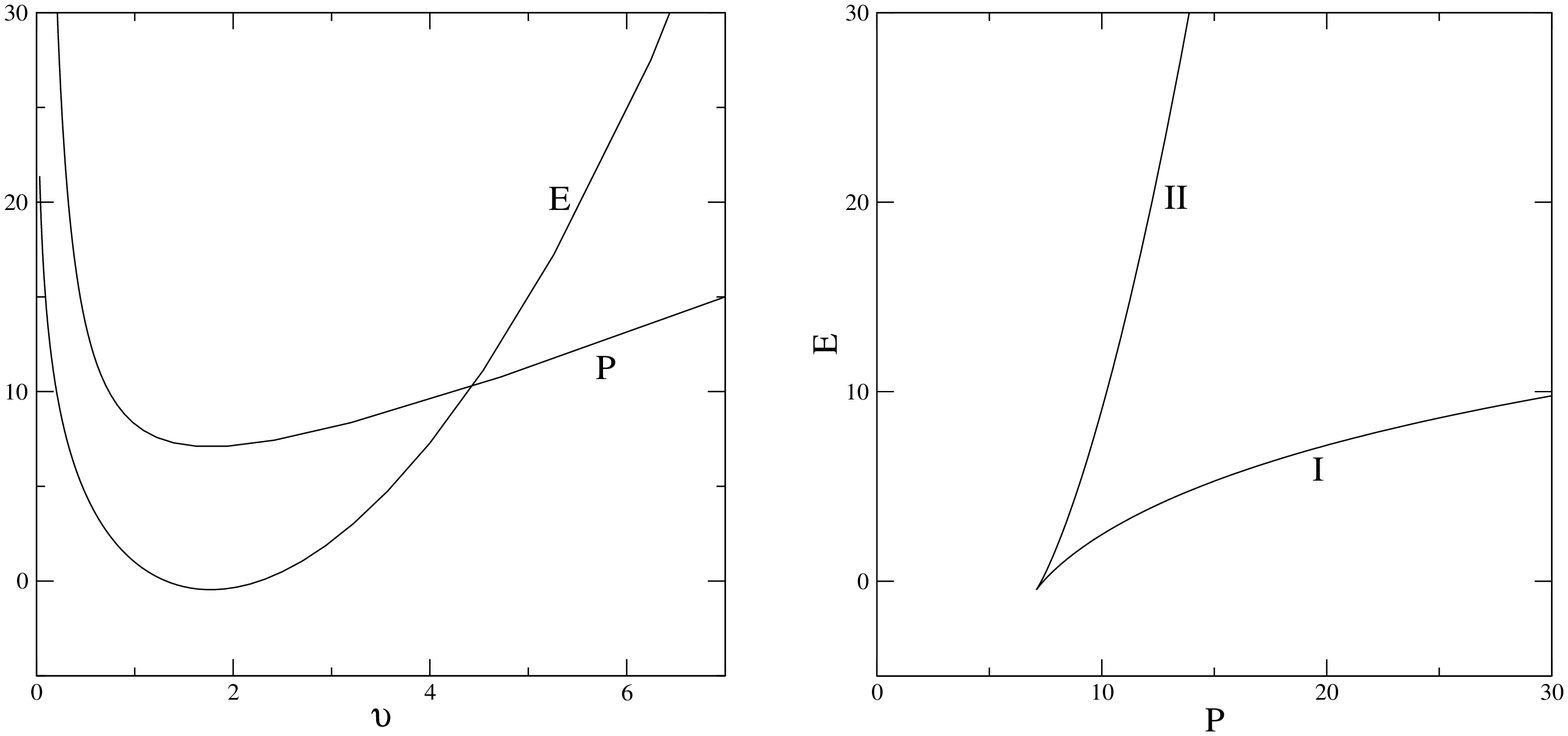,width=0.80\linewidth}
   \caption{Energy $E$ and impulse $P$ as functions of velocity $v$
(left panel) and $E$ vs $P$ dispersion (right panel)
for an electron-positron pair ($e_1=-e_2$) in Kelvin motion.
   }  \label{fig:pn_energy}
\end{figure}

For an explicit demonstration we make the special choice of potential energy
\begin{equation}  \label{eq:A10}
 V = 2\pi\, \ln|\bm{r}_1-\bm{r}_2|
\end{equation}
in order to model the behavior of VA pairs at large relative distance
\cite{pokrovskii85,kovalev02}.
For a graphical illustration we also make the special choice
of constants $m=1$ and $eB=2\pi$, as suggested by Eq.~(\ref{eq:A6}),
to write
\begin{equation}  \label{eq:A11}
 v = \frac{1}{d}\,,\quad E = \frac{1}{d^2} + 2\pi\, \ln d\,,\quad
P = \frac{2}{d} + 2\pi\,d
\end{equation}
where we note that both $E$ and $P$ acquire a minimum at a distance
$d=d_0=1/\sqrt{\pi}$ or velocity $v=v_0=\sqrt{\pi}$.
The dependence of $E$ and $P$ on the velocity $v$ as well as
the $E$ vs $P$ dispersion are shown in Fig.~\ref{fig:pn_energy}
and are found to be closely analogous to the results of
Fig.~\ref{fig:kelvin_energy}
pertaining to Kelvin motion of VA pairs.
In particular, for a widely separated pair (branch I)
we find from Eq.~(\ref{eq:A11}) that $P\sim 2\pi\,d,\; vP \sim 2\pi,\; v=1/d$,
and $E=2\pi\,\ln (P/P_0)$ with $P_0=2\pi$,
in close analogy with the asymptotic results of
Eqs.~(\ref{eq:estimates1})--(\ref{eq:kelvinpairenergy}).
The appearance of a cusp and consequently of branch II in the
spectrum is also notable.
But the details of branch II are different than those
of Fig.~\ref{fig:kelvin_energy} and Eq.~(\ref{eq:kelvinpairenergy2}).
There is now no upper limit in the velocity $v$.
In fact, all $v, E$ and $P$ in Eq.~(\ref{eq:A11}) diverge in the limit of small $d$ and
$E\sim P^2/4$, which coincides with the dispersion $P^2/2M$
of a free particle with
mass equal to the total mass of the pair $(M=2 m = 2)$.

We have also carried out a stability analysis of the special Kelvin-like
solution (\ref{eq:A6a}) to find that the motion is marginally stable along branch I
but becomes unstable along branch II.
This conclusion is in agreement with a similar result obtained in Ref.~\cite{jones86}
in the case of a vortex ring in a superfluid, as is further discussed in the concluding
remarks of our Section~\ref{sec:kelvin}.

As a last example we consider the case of 2D motion of two like charges $e_1=e_2=e$.
Then a special solution of Eq.~(\ref{eq:A1}) is given by
\begin{equation}  \label{eq:A12}
x_1 = - x_2 = R\,\cos\omega t\,,\quad y_1 = -y_2 = R\,\sin\omega t\,,\quad z_1=z_2=0
\end{equation}
which describes a pair rotating at constant radius $R=d/2$ and angular
frequency $\omega=v/R$ where the velocity $v$ is determined from the algebraic equation
\begin{equation}  \label{eq:A13}
 \frac{m v^2}{R} + e B v - V'(d) = 0
\end{equation}
that expresses the exact balance of the centrifugal, the magnetic, and the mutual force.
The conserved energy is still calculated from Eq.~(\ref{eq:A3}) with
$v_1^2=v_2^2=v^2$, but the conservation of the impulse $\bm{P}$ of Eq.~(\ref{eq:A4})
is simply equivalent to the statement that rotation takes place around a
fixed guiding center. More relevant is now the conserved angular momentum
(impulse) which is given by
\begin{equation}  \label{eq:A14}
 \am = m (x_1 \dot{y}_1 - y_1 \dot{x}_1) + \frac{e_1 B}{2} (x_1^2 + y_1^2)
   + m (x_2 \dot{y}_2 - y_2 \dot{x}_2) + \frac{e_2 B}{2} (x_2^2 + y_2^2),
\end{equation}
where the overdot denotes time derivative. Again, the angular momentum
differs from its standard mechanical expression by important
field-dependent terms.

\begin{figure}
\epsfig{file=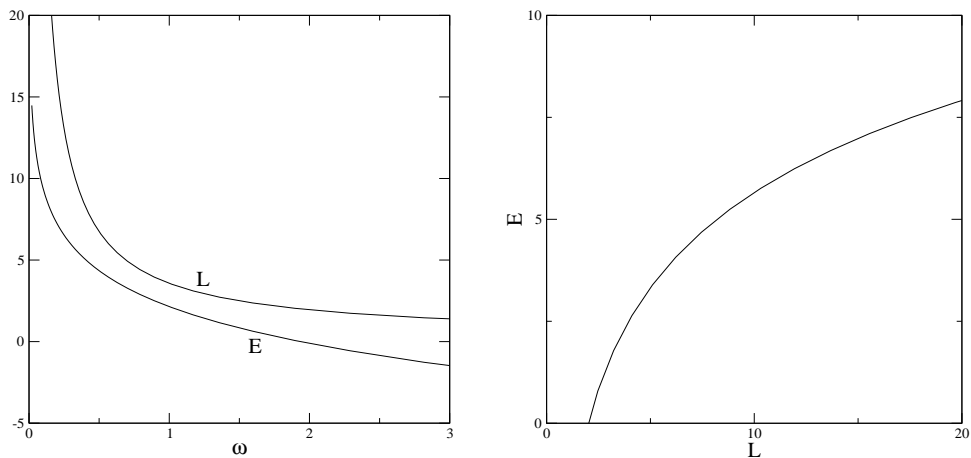,width=0.80\linewidth}
   \caption{Energy $E$ and angular impulse $\am$ as functions of
angular velocity $\omega$
(left panel) and $E$ vs $\am$ dispersion (right panel)
for a pair of like charges ($e_1=e_2$) rotating around
a fixed guiding center.
   }  \label{fig:pp_energy}
\end{figure}

Now, for the specific choice of potential energy given by Eq.~(\ref{eq:A10}),
and constants $m=1, e_1=e_2=e$ and $eB=2\pi$,
the algebraic equation (\ref{eq:A13}) yields
\begin{equation}  \label{eq:A15}
 v = \pi \left(\sqrt{R^2+\frac{1}{\pi}} - R\right)\,,
\quad R \equiv \frac{d}{2}\,,
\end{equation}
while the angular velocity $\omega$, the energy $E$, and the angular momentum $\am$,
read
\begin{equation}  \label{eq:A16}
 \omega = \frac{v}{R}\,,\quad E = v^2 + 2\pi\,\ln(2R)\,,\quad
 \am = 2 R v + 2\pi\,R^2.
\end{equation}
In view of Eq.~(\ref{eq:A15}) all three quantities in (\ref{eq:A16})
are expressed
in terms of a single parameter $R=d/2$. As a check of consistency one may
verify the relation $\omega=dE/d\am$ using Eqs.~(\ref{eq:A15}-\ref{eq:A16}).

The dependence of $E$ and $\am$ on angular velocity $\omega$ as well as the
$E$ vs $\am$ dispersion are shown in Fig.~\ref{fig:pp_energy}.
Again there exists a close analogy with the results of Section~\ref{sec:rotate}
on rotating VA pairs.
In particular, for large diameter $d$, Eqs.~(\ref{eq:A16}) yield
$\am\sim \frac{\pi}{2} d^2,\; \omega \am \sim \pi,\; \omega\sim 2/d^2$,
and $E = \pi \ln(\am/\am_0)$ with $\am_0=\pi/2$,
which should be compared with the asymptotic results for VA pairs
given in Eqs.~(\ref{eq:estimates2})--(\ref{eq:rotatepairenergy}).
On the other hand, some quantitative differences arise at small $d$, where the
angular momentum vanishes as expected ($\am\sim \sqrt{\pi} d$) but the
angular frequency diverges ($\omega \sim 2\sqrt{\pi}/d$).
Furthermore, the energy $E$ does not reach a finite value at $\am=0$
(as was the case for rotating VA pairs) but diverges logarithmically
to minus infinity.

Finally, an analysis of mechanical stability \cite{dialynas97} shows that
circular motion of two like charges in a magnetic field is marginally stable
 for all values of $d$, in contrast to the Kelvin motion discussed earlier
in this section which becomes unstable at small $d$.
However, unlike Kelvin motion which proceeds with no acceleration,
a rotating pair is expected to radiate when full electrodynamics is turned on.
Surely, this is also a source of instability and may indicate a similar instability
for the rotating VA pairs discussed in Section~\ref{sec:rotate}.

\end{appendix}

\end{document}